\documentclass[intlimits,twoside,a4paper]{article}

\usepackage[cp1251]{inputenc}

\usepackage[eqsecnum]{cmpj3}
\issue{2025}{28}{3}{33702}
\doinumber{10.5488/CMP.28.33702}

\title[Beyond conventional half-metals]%
{Beyond conventional half-metals: gapless states and  spin gapless semiconducting behavior in X\textsubscript{2}MnGa (X = Ti, Ir)  Heusler compounds\thanks{``\ldots{}I wouldn't stand
by and see the rules broken --- because right is right, and wrong
is wrong, and a  body ain't got no business doing wrong when he
ain't ignorant and knows better.'' (Twain~M., \textit{The Adventures of Huckleberry Finn}, 1884).}}
\author[N. Bouteldja, N. Hacini, I. Ouadha, H. Rached]{N. Bouteldja\orcid{0000-0002-9356-7238}\refaddr{label1}\thanks{Corresponding author: \email{n.bouteldja@univ-chlef.dz}.},
        N. Hacini\orcid{0000-0002-0837-9989}\refaddr{label2},
        I. Ouadha\orcid{0009-0000-8187-5914}\refaddr{label3},
        H. Rached\orcid{0000-0003-3867-576X}\refaddr{label3}}
\addresses{
	\addr{label1} Theoretical Physics and Materials Physics Laboratory, Hassiba Benbouali University, Chlef, Algeria\addr{label2} Condensed Matter Sciences Laboratory, Oran 1 Ahmed Ben Bella University, Oran, Algeria\addr{label3} Magnetic Materials Laboratory, Djillali Liabes University, Sidi Bel-Abbes, Algeria
}
\date{Received May 19, 2025, in final form July 11, 2025}

\begin{document}

\maketitle

\begin{abstract}
The search for high-performance spintronic materials motivates the exploration of Heusler alloys with unconventional electronic properties. Using density functional theory with Hubbard correction (DFT+$U$, $U = 4$~eV), we investigate X$_2$MnGa (X = Ti, Ir) alloys, which stabilize in the ferromagnetic L2$_1$-type structure with strong thermodynamic stability. Electronic structure calculations reveal contrasting behaviors: Ti$_2$MnGa transitions from a metallic L2$_1$-type phase to a spin gapless semiconductor (SGS) in the XA-type, while Ir$_2$MnGa exhibits gapless half-metallicity behavior in the L2$_1$-type but becomes half-metallic in the XA-type. The magnetic properties are governed by \textit{spd} hybridization between Mn-3$d$ and X-$d$/Ga-$p$ states, which stabilizes ferromagnetism and tailors electronic states near the Fermi level. The Hubbard $U$ correction proves essential for accurately describing the correlated Mn-3$d$ electrons. These alloys combine structural stability with tunable electronic and magnetic properties, offering a promising platform for spin-polarized transport in next-generation spintronic devices.
%
%
%\pacs 31.15.eg, 85.75.-d,  75.50.-y, 72.25.Ba, 72.25.Dc
\end{abstract}

\section{Introduction }

%\doclicenseThis
	
The accelerated development of spintronic systems, which exploit both the charge degree of freedom and quantum-mechanical spin of electrons, has necessitated the discovery and design of advanced functional materials with precisely tunable electronic structure and magnetic order parameters. Among the most promising candidates are Heusler alloys, a versatile class of intermetallic compounds that exhibit remarkable functionalities such as half-metallic ferromagnetism, 100\% spin polarizability and high Curie temperatures, making them ideal for spin-based electronics \cite{palmstrom2016,chatterjee2021}. These materials have found applications in magnetic random-access memory (MRAM), spin-injection devices, high-density data storage technologies, giant, tunnel and colossal magnetoresistive (GMR, TMR, CMR) sensors~\cite{graf2016,elphick2021,hirohata2022}.
	
Heusler alloys are broadly categorized into full-Heusler (X$_2$YZ) and half-Heusler (XYZ) structures, where X and Y are typically transition metals and Z is a main-group element. Their electronic structure can be precisely tuned by varying composition, leading to diverse transport behaviors \cite{galanakis2016}, including half-metallic (HM), spin gapless semiconducting (SGS) and gapless half-metallic (GHM) states \cite{gao2013,galanakis2006,skaftouros2013,wang2016}.
	
The exceptional magnetic properties of Heusler alloys stem from the strong \textit{spd} hybridization between transition metal (Mn, Co, Fe, \ldots) \textit{d}-states and main-group (Ga, Al, Si, \ldots) \textit{p}-states, which governs their electronic and magnetic behavior. This hybridization leads to the formation of spin-polarized bands near the Fermi level, stabilizing ferromagnetic (FM) or ferrimagnetic (FiM) ordering. The nature of magnetic coupling in these alloys is highly sensitive to atomic arrangement, composition and interatomic distances, allowing for precise tuning of their magnetic moments and exchange interactions \cite{zhang2015,amrich2018,ahmed2021}.
	
In full-Heusler alloys (X$_2$YZ), the magnetic moment is primarily localized on the transition metal (X, Y) sites, with Mn-based alloys often exhibiting high spin polarization due to their half-filled 3\textit{d} orbitals~\cite{sasioglu2008}. Co-based Heusler alloys (Co$_2$MnSi, Co$_2$FeAl, \ldots) are classic half-metallic ferromagnets~\cite{aguayo2011,semiannikova2021,marchenkov2014}, where one spin channel is metallic while the other is semiconducting, enabling 100\% spin-polarized currents. Ferrimagnetic Heusler alloys (Mn-based compounds) exhibit compensated magnetic moments \cite{chen2020,stinshoff2017,gavrea2020}, where antiparallel alignment of Mn spins results in a net zero magnetization, making them ideal for low-energy-loss spintronic devices.
	
Martensitic transformations in Mn-rich Heusler alloys (Ni-Mn-Ga, Ni-Mn-In, Ni-Mn-Sn, \ldots) enable remarkable multifunctional properties \cite{khan2012,zheng2013,sandeep2021}, including temperature or stress induced shape-memory effects (SME) and giant magnetocaloric effects (MCE), facilitating applications in magnetic shape-memory actuators and strain-mediated spintronics where magnetic anisotropy is controlled through lattice deformation. Beyond these conventional properties, recent discoveries have unveiled novel electronic states in Heusler alloys. SGS like Cr$_2$ZnSi \cite{chen2018} exhibit a direct touch between the valence and conduction bands at the Fermi level in one spin channel, enabling fully spin-polarized transport with minimal energy excitation, ideal for low-power spintronics. Similarly, Ti$_2$MnAl \cite{lukashev2016} shows bandgap closure in the spin-up channel, allowing near-zero-energy spin-polarized conduction for high-speed devices. Meanwhile, Mn$_3$Ga \cite{fan2020} exhibits characteristic gapless half-metallic behavior with 100\% spin polarization, featuring a unique electronic structure where one spin channel shows gapless semiconducting properties while the other remains metallic. Theoretical studies suggest that this material possesses compensated ferrimagnetic ordering and may undergo martensitic transformations, similar to the related Mn-based Heusler alloys. These Heusler compounds, with their unique electronic structures and magnetotransport properties, are pivotal for advanced spintronics.
	
In this work, we employ state-of-the-art first-principles calculations incorporating Hubbard \textit{U} corrections (DFT+\textit{U}) to conduct a comprehensive and systematic investigation of the structural, electronic and magnetic properties of X$_2$MnGa (X = Ti, Ir) Heusler alloys. Our computational approach combines density functional theory with the generalized gradient approximation (GGA-PBE) for exchange-correlation effects, augmented by a Hubbard \textit{U} correction of 4.0 eV applied to the strongly correlated 3\textit{d}-5\textit{d}-4\textit{f} states in order to accurately capture their localized character and hybridization effects.

\section{Calculation details}
To account for strong electron correlations, first-principles calculations were performed using the full-potential linearized augmented plane-wave (FPLAPW) method as implemented in the \textsc{WIEN2k} code~\cite{blaha1990,tran2006}, within the framework of density functional theory (DFT) \cite{Argaman2000}. The exchange-correlation effects were treated using the GGA-PBE \cite{perdew1996}, with an additional Hubbard $U$ correction (GGA+$U$, $U = 4.0$\,eV) applied to the Mn-$3d$ states to account for strong electron correlations~\cite{tas2022}. This approach mitigates the self-interaction error in standard GGA for localized $d$-electrons. The muffin-tin radii ($R_{\text{MT}}$) were chosen to satisfy $R_{\text{MT}} \times K_{\text{max}} = 8.0$, ensuring a well-converged basis set, while Brillouin zone integration was carried out using a $10 \times 10 \times 10$ $\mathbf{k}$-mesh (1000 $\mathbf{k}$-points). The self-consistent calculations converged with an energy criterion of $10^{-5}$\,Ry for electronic relaxation. The electronic configurations of the constituent elements were explicitly considered to accurately describe their valence states and hybridization effects in the system: Ti ($3d^2 4s^2$), Ir ($4f^{14} 5d^7 6s^2$), Mn ($3d^5 4s^2$) and Ga ($3d^{10} 4s^2 4p^1$).

\section{Spin-dependent bandscapes}
\begin{figure}[htb]
	\centering
	\includegraphics[width=0.9\linewidth]{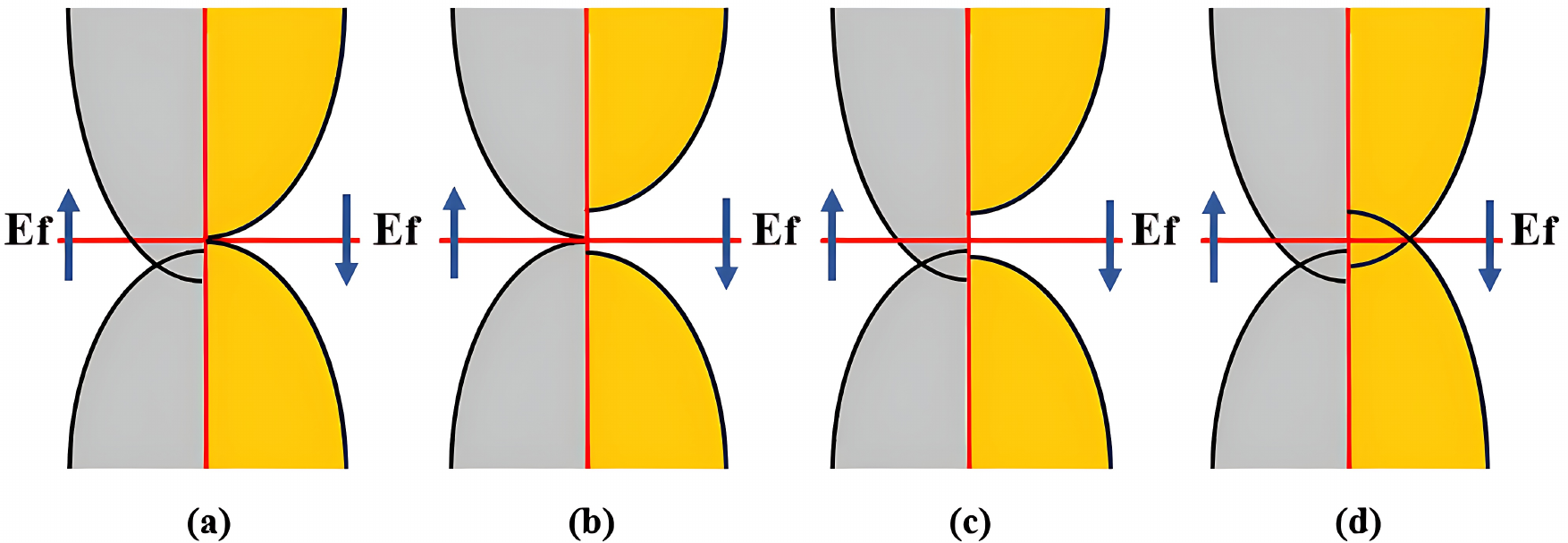}
	\caption{(Colour online) Spin-polarized band structures: (a) gapless half-metal, (b) spin gapless semiconductors, (c) half-metals, (d) magnetic metal.}
    \label{fig:spin_bands}
\end{figure}

Figure~\ref{fig:spin_bands} illustrates four distinct types of spin-dependent band structures, each enabling unique functionalities in spintronics: (a) GHM exhibit Dirac-like metallic spin-$\uparrow$ states with linear dispersion, while spin-$\downarrow$ remains semiconducting with zero bandgap, enabling high-speed spintronic devices~\cite{Wang2018,Wang2020}; (b) SGS have gapless spin-$\uparrow$ bands (zero DOS at $E_{\rm{F}}$) while spin-$\downarrow$ bands are gapped, enabling ultra-low-power spin transistors and quantum sensors \cite{Liu2017}; (c) HM are 100\% spin-polarized ($\uparrow$=metallic, $\downarrow$=semiconducting), ideal for MRAM and spin injection \cite{Belashchenko2012,Shaughnessy2011}; (d) magnetic metals (MM) conduct both spin-$\uparrow$ and spin-$\downarrow$ electrons with partial spin polarization, making them useful for spin valves and GMR sensors \cite{Jedema2002}.

\section{Results and discussion}

\subsection{Structural properties}

\begin{figure}[h]
	\centering
	\includegraphics[width=0.52\linewidth]{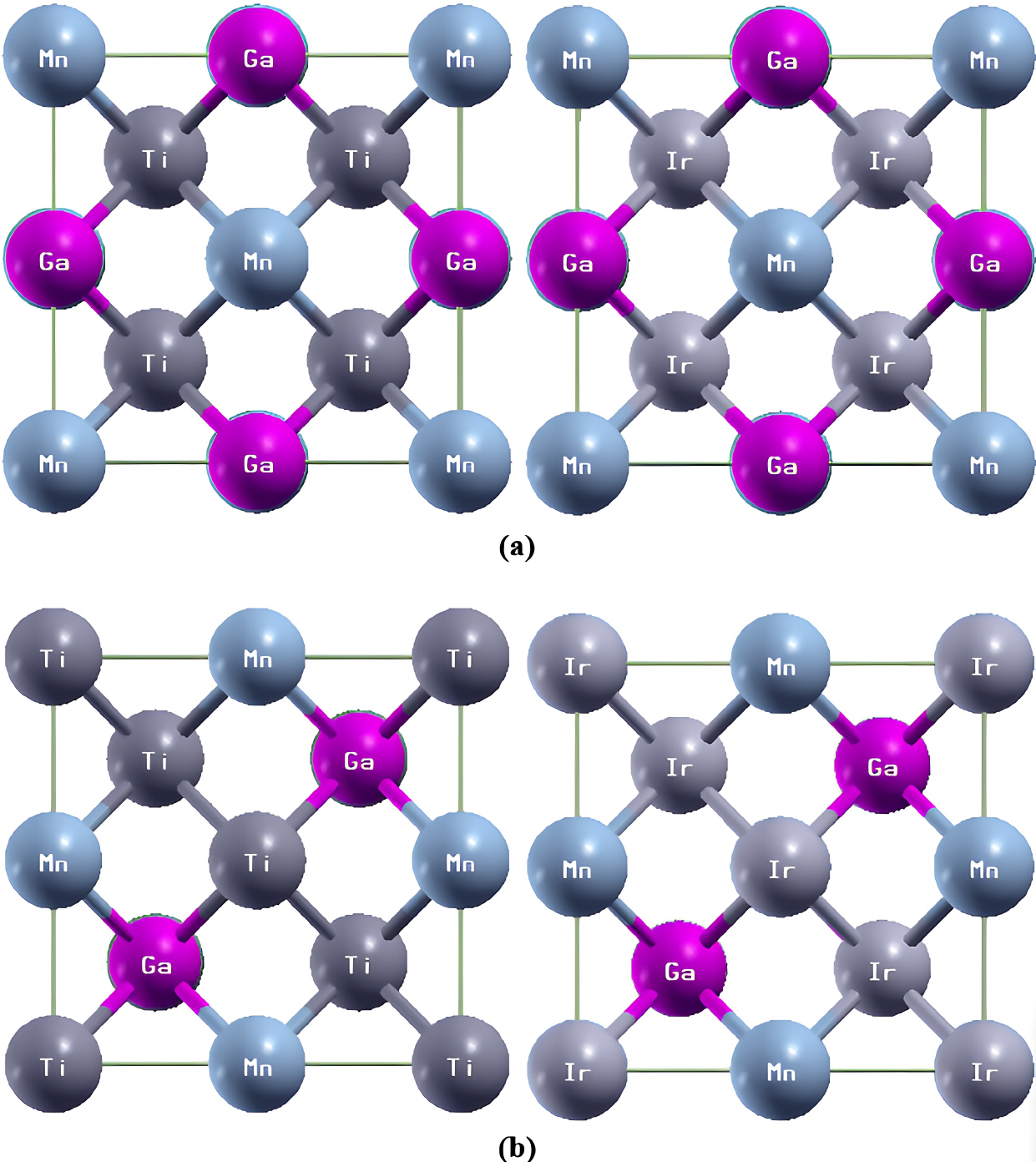}
	\caption{(Colour online) Crystal structures of Ti\textsubscript{2}MnGa and Ir\textsubscript{2}MnGa: (a) conventional L2\textsubscript{1}-type structure (space group Fm$\overline{3}$m) and (b) inverse XA-type structure (space group F$\overline{4}$3m). }
	\label{fig:crystal_tructures }
\end{figure}

{The crystal structures of the full Heusler alloy X$_2$MnGa (X = Ti, Ir) are shown in figure~\ref{fig:crystal_tructures }, displaying both: (a) the conventional L2$_1$-type structure (space group Fm$\overline{3}$m, No. 225), (b) the inverse Heusler XA-type structure (space group F$\overline{4}$3m), where both structures consist of four interpenetrating face-centered cubic (fcc) sublattices with distinct atomic arrangements: in the L2$_1$-type structure, the Ti/Ir atoms occupy the (0.25, 0.25, 0.25) and (0.75, 0.75, 0.75) Wyckoff positions, Mn occupies (0, 0, 0) and Ga is located at (0.5, 0.5, 0.5), while for the inverse XA-type structure, one Ti/Ir atom sits at (0, 0, 0), another at (0.25, 0.25, 0.25), Mn at (0.5, 0.5, 0.5) and Ga at (0.75, 0.75, 0.75), with muffin-tin radii ($R_{\mathrm{MT}}$) typically chosen as $\rm{Ti} = 2.2$, $\rm{Ir} = 2.5$, $\rm{Mn} = 2.0$ and $\rm{Ga} = 2.1$ atomic units to ensure a proper overlap in the DFT calculations while minimizing computational artifacts.}

\begin{figure}[h]
	\centering
	\includegraphics[width=0.9\linewidth]{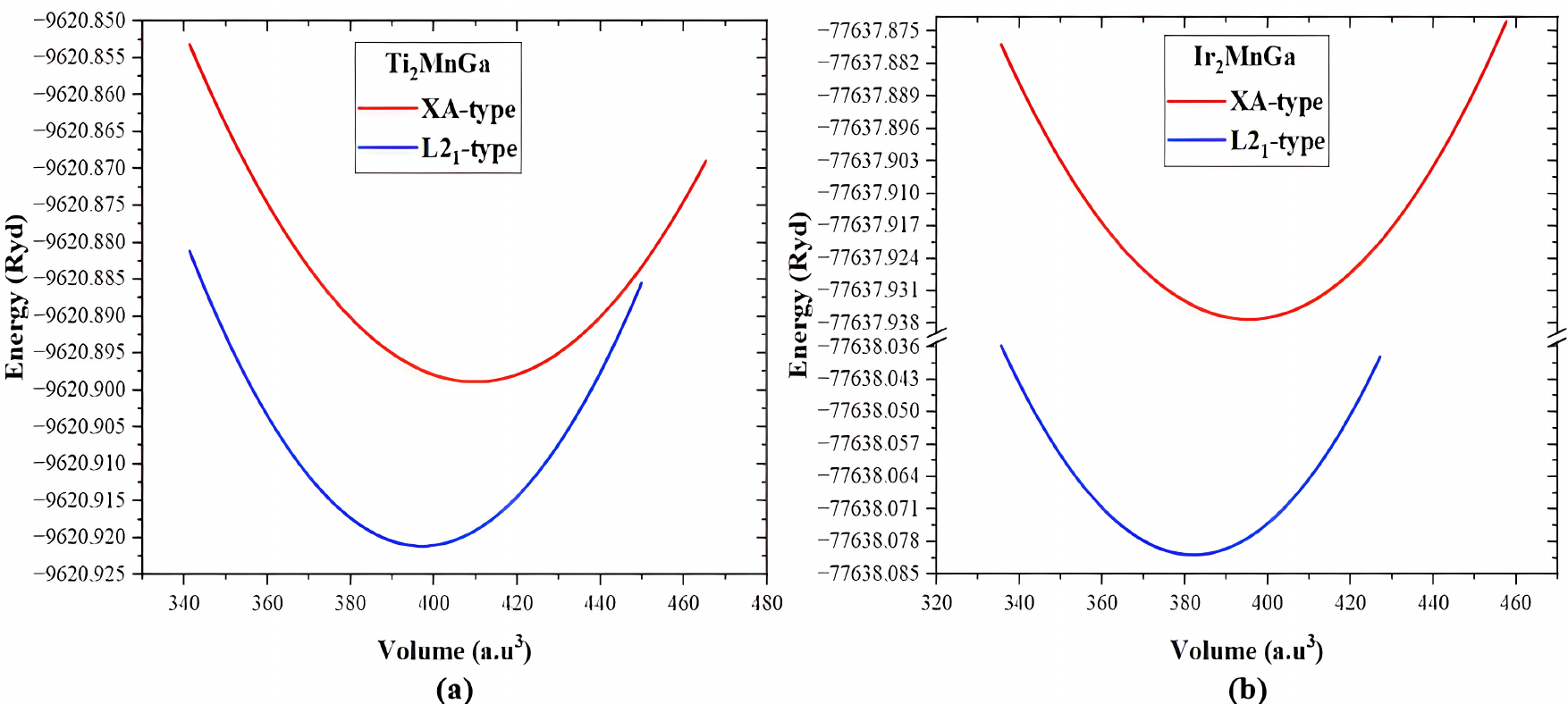}
	\caption{(Colour online) Energy versus volume curves for (a) Ti\textsubscript{2}MnGa and (b) Ir\textsubscript{2}MnGa in both L2\textsubscript{1}-type and XA-type structures, calculated using PBE+$U$ ($U = 4$\,eV).}
	\label{fig:energy_volume}
\end{figure}

To determine the stable ground-state configurations, we performed total energy versus volume ($E$--$V$) calculations for both the $L2_1$-type and XA-type structures, then fitted the data using the Birch--Murnaghan equation of state \cite{murnaghan1944}.
\begin{equation}
	E(V) = E_0 + \frac{B}{B' (B' + 1)} \left[ V \left( \frac{V_0}{V} \right)^{\beta} - V_0 \right] + \frac{B}{B'} (V - 1).
\end{equation}
{Figure~\ref{fig:energy_volume} compares the total energies of: (a) Ti$_2$MnGa, (b) Ir$_2$MnGa in L2$_1$-type (Fm$\overline{3}$m) and XA-type (F$\overline{4}$3m) configurations. Our calculations confirm that the L2$_1$-type structure is energetically favorable, exhibiting a lower total energy than the XA-type structure. This stabilization primarily originates from optimized chemical bonding and hybridization effects between the constituent elements. In the L2$_1$ structure, the Mn atoms occupy a high-symmetry position that maximizes favorable $d$--$d$ orbital overlap with neighboring transition metals (Ti/Ir) while maintaining optimum charge transfer to Ga $sp$-states. This coordination geometry enhances covalent bonding and band filling efficiency compared to the XA-type structure, resulting in greater thermodynamic stability.}

Table~\ref{tab:structural_properties} presents the computed structural parameters of X$_2$MnGa (X = Ti, Ir). The lattice constants~($a$) were optimized through energy-volume calculations and the bulk modulus~($B$) along with its pressure derivative~($B'$) were extracted from Birch--Murnaghan equation of state fitting. The corresponding equilibrium energy~($E_0$) at the minimum energy configuration is also reported.

The calculated structural parameters reveal several key trends. The L2$_1$-type phase consistently exhibits smaller lattice parameters ($a$) compared to the XA-type structure for both compounds, indicating denser atomic packing. Bulk modulus ($B$) values are considerably higher in the L2$_1$-type phase (148.99~GPa for Ti$_2$MnGa compared with 236.17~GPa for Ir$_2$MnGa), demonstrating greater structural rigidity. The pressure derivative ($B'$) values between 3.77--5.64 suggest normal elastic behavior for all configurations. The lower minimum energies ($E_{\text{min}}$) of the L2$_1$-type phase confirm its thermodynamic stability over the XA-type structure, with energy differences of 0.0228~Ry (Ti$_2$MnGa) and 0.1432~Ry (Ir$_2$MnGa), respectively. These results correlate with the stronger $d$--$d$ hybridization in the L2$_1$ configuration observed in our electronic structure analysis.

\begin{table}[h]
	\caption{Lattice parameters and properties of Ti$_2$MnGa and Ir$_2$MnGa compounds calculated via PBE+$U$.}
	\label{tab:structural_properties}
	\centering
	\begin{tabular}{|l||c|c||c|c|}
		\hline
		\small\textbf{Property} & \multicolumn{2}{c||}{\small\textbf{L2$_1$ (Fm$\overline{3}$m)}} & \multicolumn{2}{c|}{\small\textbf{XA (F$\overline{4}$3m)}} \\
		\cline{2-5}
		& \small\textbf{Ti$_2$MnGa} & \small\textbf{Ir$_2$MnGa} & \small\textbf{Ti$_2$MnGa} & \small\textbf{Ir$_2$MnGa} \\
		\hline\hline
		\small\textbf{Lattice constant} & & & & \\
		\small$a$ (\AA) &\small 5.9138 & \small5.8461 &\small 5.9815 &\small 5.9036 \\
		\hline
		\small\textbf{Bulk modulus} & & & & \\
		\small$B$ (GPa) & \small 148.9958 &\small 236.1689 &\small 108.9095 &\small 199.1957 \\
		\hline
		\small\textbf{Pressure derivative} & & & & \\
		\small$B'$ &\small 5.6251 &\small 5.1368 &\small 3.7698 &\small 4.6396 \\
		\hline
		\small\textbf{Minimum energy} & & & & \\
		\small$E_{\text{min}}$ (Ry) &\small $-9620.921074$ &\small $-77638.080872$ &\small $-9620.898279$ &\small $-77637.937667$ \\
		\hline
	\end{tabular}
\end{table}

\subsection{Electronic properties}

\subsubsection{Band structure}

\begin{figure}
	\centering
	\includegraphics[width=1\linewidth]{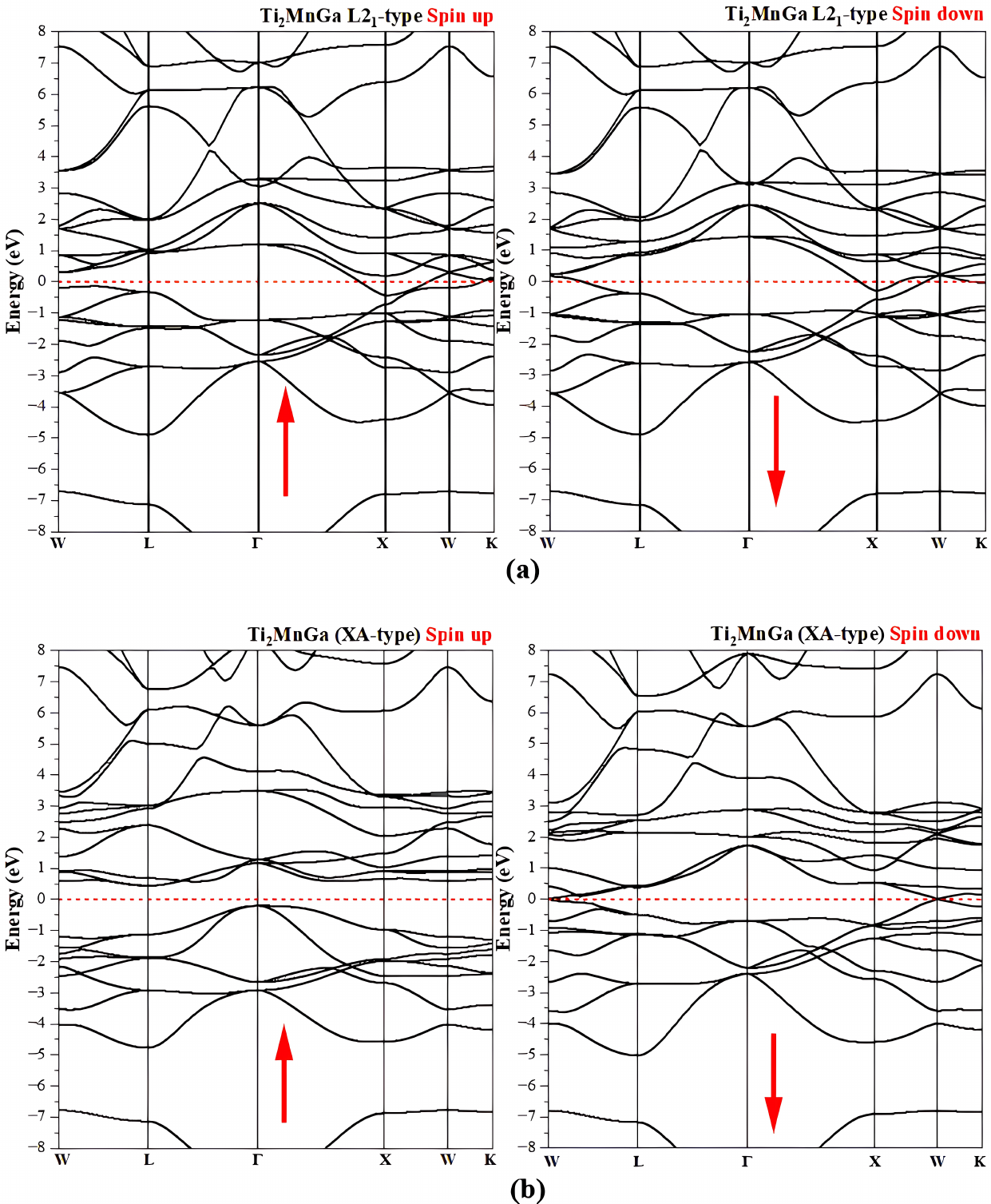}
    \caption{(Colour online) Spin-polarized band structure of Ti\textsubscript{2}MnGa calculated using PBE+$U$ ($U = 4$\,eV) for both (a) L2\textsubscript{1}-type and (b) XA-type structures. }
    \label{fig:band_structures 4}
\end{figure}

\begin{figure}
	\centering
	\includegraphics[width=1\linewidth]{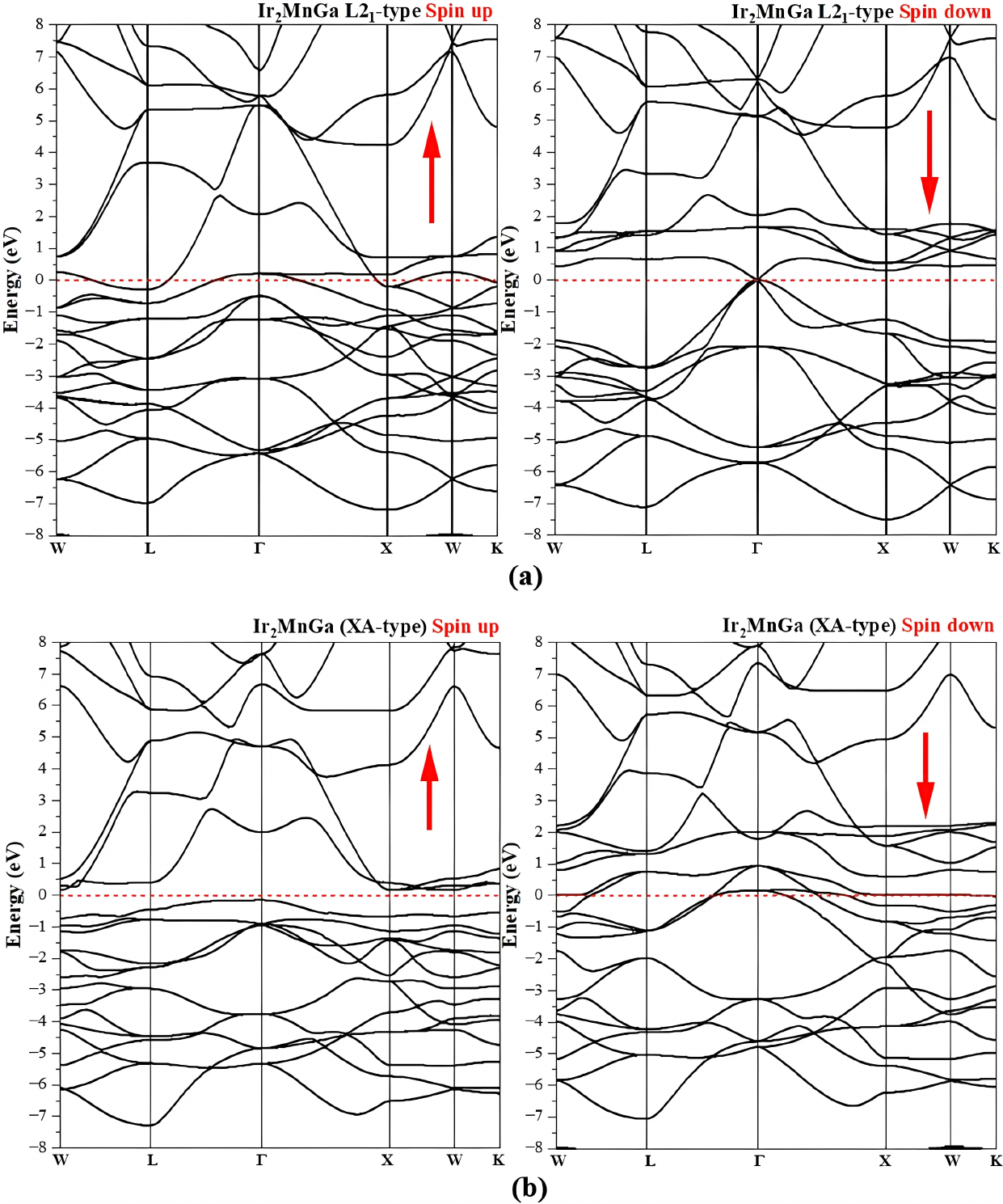}
    \caption{(Colour online) Spin-polarized band structure of Ir\textsubscript{2}MnGa calculated using PBE+$U$ ($U = 4$\,eV) for both (a) L2\textsubscript{1}-type  and (b) XA-type structures.}
\label{fig:band_structures 5}
\end{figure}

The electronic band structures of X$_2$MnGa (X = Ti, Ir) Heusler alloys reveal fundamentally distinct spin-dependent behaviors, as shown in figures~\ref{fig:band_structures 4} and~\ref{fig:band_structures 5}. These differences originate from their unique electronic configurations, which dramatically influence their spin-polarized transport properties.

Ti\textsubscript{2}MnGa in the XA-type structure exhibits a characteristic spin gapless semiconductor behavior (confirmed by \cite{Goraus2019}). As visible in Figure~\ref{fig:band_structures 4}b, the spin-$\uparrow$ channel presents an indirect band gap of 0.528~eV between the $\Gamma$ and L points, while the spin-$\downarrow$ channel forms a zero-gap configuration, where the valence and conduction bands touch precisely at the Fermi level ($E_F$). This unique band alignment enables efficient spin-polarized transport, as charge carriers in the spin-$\downarrow$ channel can be excited without overcoming an energy barrier, while the spin-$\uparrow$ channel remains semiconducting. When transformed to the L2\textsubscript{1}-type structure (figure~\ref{fig:band_structures 4}a), this delicate SGS configuration is lost. Both spin channels cross $E_{\rm{F}}$, resulting in a conventional ferromagnetic metallic state. The disappearance of the SGS behavior in the L2\textsubscript{1} phase can be attributed to the altered crystal field splitting and hybridization between the Ti-$3d$, Mn-$3d$, and Ga-$4p$ orbitals, which modifies the band structure near $E_{\rm{F}}$. This transition underscores the critical role of atomic ordering in determining the electronic properties.

Ir\textsubscript{2}MnGa demonstrates markedly different electronic behavior compared to its Ti counterpart, as evidenced in figure~\ref{fig:band_structures 5}a. The strong spin-orbit coupling (SOC) from iridium $5d$ electrons plays a decisive role in shaping its electronic structure. In the L2\textsubscript{1}-type phase, the system exhibits a rare gapless half-metallic state, the spin-$\uparrow$ channel shows metallic conductivity with bands crossing the Fermi level~($E_{\rm{F}}$), while the spin-$\downarrow$ channel forms a zero-gap configuration where the valence and conduction bands touch at $E_{\rm{F}}$. This unique state arises from the interplay between SOC-induced band splitting and the specific symmetry of the L2\textsubscript{1} lattice, which preserves degeneracies at certain high-symmetry points. When transformed to the XA-type structure (figure~\ref{fig:band_structures 5}b), Ir\textsubscript{2}MnGa transitions to a standard half-metal (confirmed by~\cite{Balakrishnan2022}). The spin-$\uparrow$ channel opens an indirect band gap of 0.212 eV between the $\Gamma$ and X points, while the spin-$\downarrow$ channel becomes metallic with band overlap at $E_{\text{F}}$. This change stems from the XA structure lower symmetry, which breaks degeneracies and enhances exchange splitting compared to the L2\textsubscript{1} phase. The result is a clear separation between valence and conduction bands in the minority spin channel while maintaining metallic conductivity in the majority spins.

These results demonstrate how electronic properties are intimately connected to both crystal structure and elemental composition. The Ti system shows particular phase sensitivity, while the Ir compound maintains robust spin polarization across structures. The comparative analysis reveals the complex interplay between spin-orbit coupling and crystal symmetry in these technologically important materials.

\subsubsection{Density of states }
\begin{figure}[htb]
	\centering
	\includegraphics[width=0.9\linewidth]{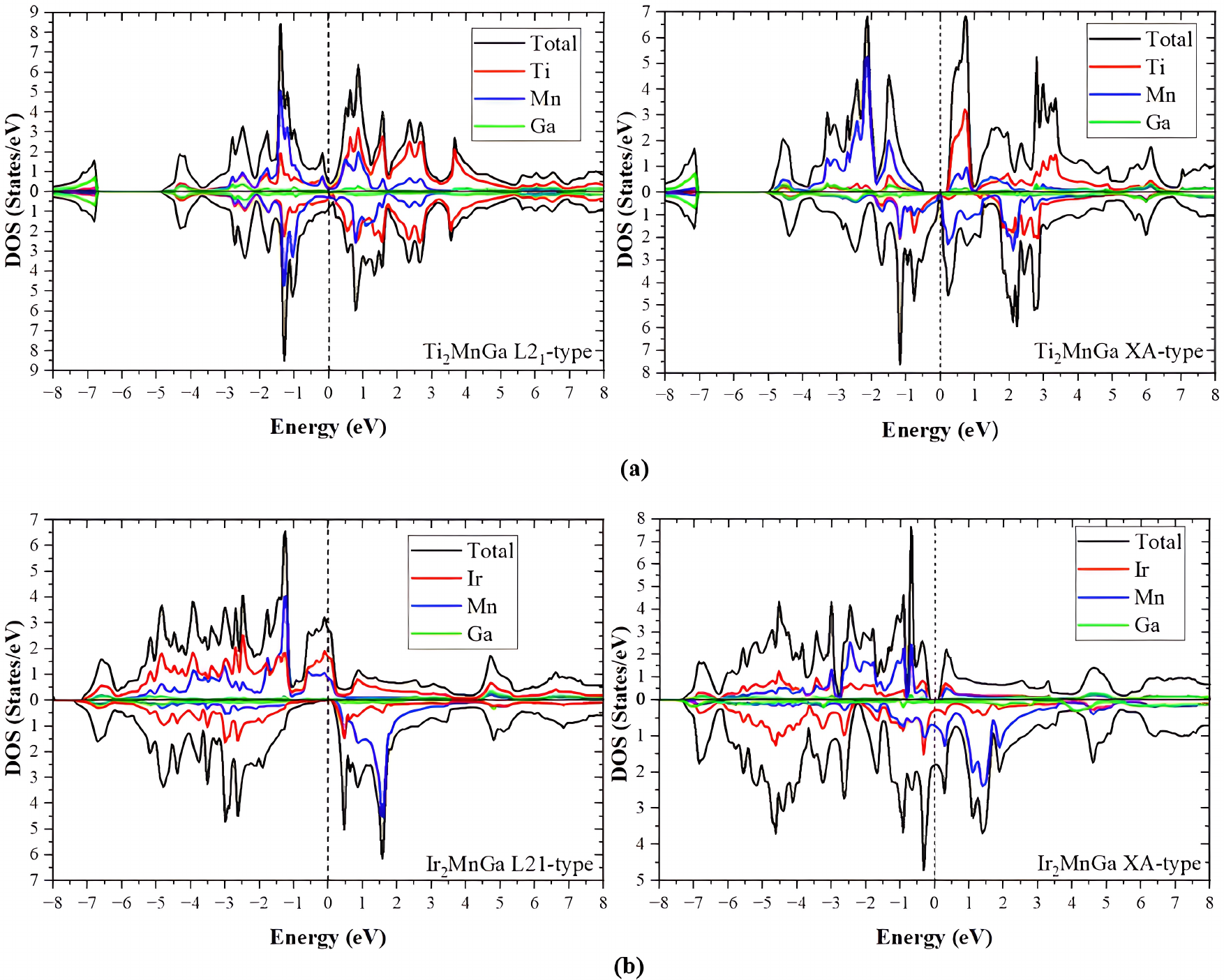}
    \caption{(Colour online) Spin-dependent total density of states for (a)~Ti\textsubscript{2}MnGa and (b)~Ir\textsubscript{2}MnGa calculated using PBE+$U$ ($U = 4$\,eV) in both L2\textsubscript{1}-type and XA-type  structures.}
    \label{fig:spin_dos}
\end{figure}

Figure~\ref{fig:spin_dos} illustrates the spin-dependent total and atomic density of states (DOS) for Ti\textsubscript{2}MnGa and Ir\textsubscript{2}MnGa in both L2\textsubscript{1}-type and XA-type structural configurations.

For Ti$_2$MnGa, while the Ti contributions are more prominent in the conduction band (0 to 5 eV), there is a noticeable overlap with Mn in the upper valence band ($-$5 to 0 eV), suggesting a degree of covalent bonding due to hybridization between Ti and Mn atoms. This hybridization is particularly strong in the L2$_1$-type structure, where the symmetric arrangement of atoms maximizes orbital overlap. The Ga contributions, while less dominant, play a crucial role in mediating charge transfer between transition metal sites (Ti and Mn), particularly in the lower valence region ($-8$ to $-7$ eV). 

For Ir$_2$MnGa, the Ir contributions are mainly concentrated around the Fermi level in the L2$_1$-type structure, showing strong overlap with Mn states in the range ($-4$ to 2\,eV). This indicates significant hybridization, which promotes metallic behavior through enhanced electron delocalization. In contrast, the XA-type structure exhibits more localized Mn states near the Fermi level and weaker Ir contributions, reflecting reduced hybridization due to the lower symmetry. These differences highlight the sensitivity of the electronic structure to the atomic arrangement and symmetry.

The superior stability and metallic character of the L2$_1$-type structure arise from its enhanced hybridization and symmetric coordination, which promote electron delocalization, as observed in Ti$_2$MnGa. In contrast, the XA-type structure exhibits anisotropic hybridization, leading to more localized states and diverse electronic behavior, such as the half-metallic properties observed in Ir$_2$MnGa, including the possible formation of a gap in certain spin channels. These differences highlight the significant influence of crystal symmetry and atomic arrangement on the electronic properties of Heusler alloys, as confirmed by our calculations.

\subsection{Magnetic properties}

The magnetic moments of X$_2$MnGa (X = Ti, Ir) in L2$_1$ and XA structures were calculated via PBE+$U$ ($U = 4$\,eV). Table~\ref{tab:magnetic} shows the atom-resolved and total moments, revealing the impact of atomic species and crystal symmetry. Comparison with the Slater--Pauling rule~\cite{slater1936,pauling1938} highlights deviations due to electronic localization and hybridization effects.  

\begin{table}[htb]
	\caption{Magnetic moments (in $\mu_{\rm B}$) of Ti$_2$MnGa and Ir$_2$MnGa compounds calculated via PBE+$U$.}
	\label{tab:magnetic}
	\centering
	\begin{tabular}{|l||c|c||c|c|}
		\hline
		\small\textbf{Property} & \multicolumn{2}{c||}{\small\textbf{L2$_1$ (Fm$\overline{3}$m)}} & \multicolumn{2}{c|}{\small\textbf{XA (F$\overline{4}$3m)}} \\
		\cline{2-5}
		& \small\textbf{Ti$_2$MnGa} & \small\textbf{Ir$_2$MnGa} &\small \textbf{Ti$_2$MnGa} &\small \textbf{Ir$_2$MnGa} \\
		\hline\hline
		\small\textbf{Interstitial moment} & & & & \\
		\small$M_{\rm int}$ &\small $-0.30669$ & \small 0.08026 &\small $-0.75716$ &\small 0.04452 \\
		\hline
		\small\textbf{Transition metal moment} & & & & \\
		\small$M_{\rm X}$ (X = Ti/Ir) &\small $-0.23601$ &\small 0.12941 &\small $-0.28886$ &\small $-0.24595$ \\
		\hline
		\small\textbf{Manganese moment} & & & & \\
		\small$M_{\rm Mn}$ &\small 3.34682 &\small 3.39502 &\small 3.15768 &\small 3.44853 \\
		\hline
		\small\textbf{Gallium moment} & & & & \\
		\small$M_{\rm Ga}$ &\small $-0.02189$ &\small 0.00140 &\small $-0.00306$ &\small $-0.02329$ \\
		\hline\hline
		\small\textbf{Total magnetic moment} & & & & \\
		\small$M_{\rm tot}$ &\small 3.49026 &\small 3.99432 &\small 1.81973 &\small 2.97786 \\
		\hline
	\end{tabular}
\end{table}

As shown in table~\ref{tab:magnetic}, Mn exhibits the largest magnetic moment (3.16--3.45$\mu_{\mathrm{B}}$), considerably reduced from the expected 5$\mu_{\mathrm{B}}$ for isolated Mn$^{2+}$. This reduction stems from three factors: strong Mn-$3d$/X-$d$/Ga-$p$ hybridization, $d$-electron delocalization and crystal field effects.

Our calculated moments agree well with literature values: 3.99$\mu_{\mathrm{B}}$ for Ir$_2$MnGa compared to the reported 4.02$\mu_{\mathrm{B}}$~\cite{Krishnaveni2019} and 3.49$\mu_{\mathrm{B}}$ for Ti$_2$MnGa relative to 3.38$\mu_{\mathrm{B}}$~\cite{Jia2014}, with minor differences attributable to computational parameters. The magnetic configuration shows antiferromagnetic Ti-Mn coupling (reducing the net moment) and structure-dependent Ir-Mn alignment.

Moment magnitudes vary with structure: the L2$_1$ phase shows higher moments (3.5--4.0$\mu_{\mathrm{B}}$) due to parallel alignment, while the XA phase exhibits reduced moments (1.8--3.0$\mu_{\mathrm{B}}$) from enhanced antiferromagnetic coupling. These results demonstrate the combined influence of spin alignment, local symmetry and electronic hybridization, validating our computational approach while revealing the parameter sensitivity.

\section{Conclusion}

This study reveals the fundamental structure-property relationships in Ti$_2$MnGa and Ir$_2$MnGa Heusler alloys through first-principles calculations. Three key findings emerge, the electronic ground state is highly sensitive to both composition and crystal symmetry, with Ti$_2$MnGa transitioning from spin gapless semiconductor (SGS) behavior in the XA-type structure (zero spin-$\uparrow$ gap and 0.5~eV spin-$\downarrow$ gap) to conventional ferromagnetic metallicity in the L2$_1$-type phase, while Ir$_2$MnGa shifts from gapless half-metallicity (L2$_1$) to conventional half-metallicity (XA-type). Magnetic properties are dominated by Mn-$3d$ states with moments reduced to 3.16--3.45$\mu_{\mathrm{B}}$ from the atomic 5$\mu_{\mathrm{B}}$ limit due to strong $d$--$d$/$d$--$p$ hybridization and crystal field effects. The Slater--Pauling rule partially governs the total moments, with deviations revealing complex interatomic exchange mechanisms. These results demonstrate how targeted substitutions (Ti, Ir) and structural control (L2$_1$, XA) can tune these alloys for specific spintronic applications requiring either fully spin-polarized transport (Ir-rich compositions) or low-power excitations (Ti-based SGS states).

\section*{Acknowledgements}

The authors would like to express their sincere gratitude to the directors and staff of the respective research laboratories for their continuous support and for providing the necessary computational facilities and resources that made this work possible.

\ukrainianpart

\title{Поза межами звичайних напівметалів: безщілинні стани та безщілинна спінова напівпровідникова поведінка в 
	сполуках Гейслера X\textsubscript{2}MnGa (X = Ti, Ir)}
\author{Н. Бутельджа\refaddr{label1},
	Н. Хачіні\refaddr{label2},
	І. Уадха\refaddr{label3},
	Х. Рашед\refaddr{label3}}
\addresses{
	\addr{label1} Лабораторія теоретичної фізики та фізики матеріалів, Університет Хассіба Бенбуалі, Шлеф, Алжир
	\addr{label2} Лабораторія конденсованих систем, Оран 1 Університет Ахмеда Бен Белли, Оран, Алжир
	\addr{label3} Лабораторія магнітних матеріалів, Університет Джиллалі Ліабес, Сіді-Бель-Аббес, Алжир
}
%
%% якщо автор є один або автори є з однієї установи:
%
%  \author{1й Автор, 2й Автор, \ldots}
%  \address{Інститут\ldots}
%
%%

\makeukrtitle

\begin{abstract}
	\tolerance=3000%
	Пошук високопродуктивних спінтронних матеріалів мотивує дослідження сплавів Гейслера з нетрадиційними електронними властивостями. Використовуючи теорію функціоналу густини з поправкою Хаббарда (DFT+$U$, $U = 4$~eV), ми досліджуємо сплави X$_2$MnGa (X = Ti, Ir), які стабілізуються у феромагнітній структурі типу L2$_1$ з сильною термодинамічною стійкістю. Розрахунки виявляють різну поведінку електронної структури: Ti$_2$MnGa переходить з металевої фази типу L2$_1$ у спін-безщілинний напівпровідник (SGS) типу XA, тоді як Ir$_2$MnGa демонструє безщілинну напівметалеву поведінку у типі L2$_1$, але стає напівметалевим у типі XA. Магнітні властивості визначаються \textit{spd} гібридизацією між станами Mn-3$d$ та X-$d$/Ga-$p$, яка стабілізує феромагнетизм та адаптує електронні стани поблизу рівня Фермі. Поправка Хаббарда $U$ виявляється важливою для точного опису скорельованих електронів Mn-3$d$. Ці сплави поєднують структурну стабільність з регульованими електронними та магнітними властивостями, створюючи перспективну основу для спін-поляризованого переносу в спінтроннічних пристроях наступного покоління.
	\keywords сплави Гейслера, спінтроніка, DFT+U, спінові безщілинні напівпровідники, напівметали, магнітні матеріали
	
\end{abstract}

\lastpage
\end{document}